\input harvmac
\def\a{{\alpha}}
\def\b{{\beta}}
\def\g{{\gamma}}

\def\d{{\delta}}
\def\s{{\sigma}}

\def\l{{\lambda}}
\def\G{{\Gamma}}

\Title
{\vbox{\hbox{IFT-P.018/98,}}
       \vbox{\hbox{HUTP-98/A040}}     }
{\vbox{\centerline{Type IIB $R^4 H^{4g-4}$ Conjectures}}}
\vskip .2in

\centerline{ Nathan Berkovits}
\vskip .2in
\centerline{Instituto de F\'{\i}sica Te\'orica, Univ. Estadual Paulista}
\centerline{Rua Pamplona 145, S\~ao Paulo, SP 01405-900, Brasil}
\vskip .3in
\centerline{ Cumrun Vafa}
\vskip .2in \centerline{Lyman Laboratory of Physics, Harvard University}
\centerline{Cambridge, MA 02138, USA}
\vskip .3in

We propose $SL(2,Z)$ (and $SL(3,Z)$) invariant
conjectures for all $R^4H^{4g-4}$ couplings
of Type IIB strings on $R^{10}$ (and $R^{8}\times T^2$),
generalizing conjectures of Green and Gutperle (and Kiritsis and
Pioline) for the $R^4$ coupling. A strong check for our conjectures
is that on $T^2$ at weak coupling, they reproduce the multiloop
scattering amplitudes which had been previously computed using
$N=2$ strings in the $N=4$ topological formalism.
Applications to $(p,q)$ string production in a background
$H$ field, generalizing Schwinger's computation
for pair production in constant $F$ field, are suggested.

\Date{3/98}

\newsec{Introduction}

Non-trivial
string duality conjectures often lead to
perturbative predictions for certain amplitudes.  Such
amplitudes are usually very special and receive corrections only
at specific genera.  A well known case of this is the $R^4$ coupling
in Type IIB theory
 which has been argued to only receive perturbative corrections
at tree-level and one-loop \ref\gross{D.J. Gross and E. Witten,
{\it Superstring Modifications of Einstein's Equations},
Nucl. Phys. B277 (1986) 1\semi N. Sakai and Y. Tanii,
Nucl. Phys. B287 (1987) 457.}\ref
\nonren{N. Berkovits, {\it Construction of $R^4$ Terms in
N=2 D=8 Superspace}, Nucl. Phys. B514 (1998) 191, hep-th/9709116}\foot
{It should be noted that there is a
possible contradiction in the literature \ref\Ien{R. Jengo and C.-J. Zhu,
{\it Two Loop Computation of the Four Particle Amplitude in
Heterotic String Theory}, Phys. Lett. B212 (1988) 313.}\
concerning the two-loop $R^4$
contribution.}.
Green and Gutperle conjectured that the $R^4$
term appears in the effective action multiplied
by the manifestly SL(2,Z)-invariant Eisenstein function
$E_{3/2} (\tau)$\ref\Green{M.B. Green and M. Gutperle,
{\it Effects of D-instantons}, Nucl. Phys. B498 (1997) 195,
hep-th/9701093.}, and their
conjecture is supported by various types of evidence\ref\Taylor
{I. Antoniadis, B. Pioline and T.R. Taylor,
{\it Calculable $\exp(1/\l)$ Effects},
hep-th/9707222.}\ref\Greentwo{M.B. Green and P. Vanhove,
{\it D-Instantons, Strings and M-theory}, Phys. Lett. B408 (1997) 122,
hep-th/9704145\semi M.B. Green and M. Gutperle and P. Vanhove,
{\it One Loop in Eleven Dimensions}, Phys. Lett. B409 (1997) 177,
hep-th/9706175.}\ref\Kiritsis{E. Kiritsis and B. Pioline,
{\it On $R^4$ Threshold Corrections in IIB String Theory and (p,q)
String Instantons}, Nucl. Phys. B508 (1997) 509, hep-th/9707018.},
in particular by the match with the genus 0 and genus 1 amplitudes
which are explicitly known.
The success of the $R^4$ conjecture naturally leads one to look
for generalizations\ref\tseytlin{J.G. Russo and A.A. Tseytlin,
{\it One Loop Four Graviton Amplitude in Eleven-Dimensional
Supergravity}, Nucl. Phys. B508 (1997) 245, hep-th/9707134.}\ref\russo
{J. Russo, {\it An Ansatz for a Non-perturbative Four Graviton
Amplitude in Type IIB Superstring Theory}, hep-th/9707241.
}, but in the absence of explicit multiloop
calculations, it is difficult to choose between different proposals.
Also, it is not apriori clear what kinds of amplitudes one
should concentrate on.

More than three year ago, we showed that $R^4 H^{4g-4}$
(or $R^4 F^{4g-4}$) terms can be computed at genus $g$
for the Type IIB (or Type IIA) superstring compactified to six
dimensions on any hyper-Kahler manifold \ref\us{N. Berkovits and
C. Vafa,
{\it N=4 Topological Strings},
Nucl. Phys. B433 (1995) 123, hep-th 9407190.  }.
Like the better known four-dimensional $R^2 F^{2g-2}$ terms \ref\topol
{M. Bershadsky, S. Cecotti, H. Ooguri and C. Vafa,
{\it Holomorphic Anomalies in Topological Field Theories}, Nucl. Phys. B405
(1993) 279, hep-th/9302103\semi
I. Antoniadis, E. Gava, K.S. Narain and
T.R. Taylor, {\it Topological Amplitudes in String Theory},
Nucl. Phys. B413 (1994) 162, hep-th/9307158\semi
M. Bershadsky, S. Cecotti, H. Ooguri and C. Vafa,
{\it
Kodaira-Spencer Theory of Gravity and Exact Results for
Quantum String Amplitudes}, Comm. Math. Phys. 165 (1994) 311,
hep-th/9309140.},
these six-dimensional terms can be expressed in terms of ($N=4$) topological
string computations, which are in turn equivalent to the partition
function of the $N=2$ string on the corresponding four manifold.
The topological reformulation of the amplitudes allows one to find
methods to compute them explicitly, as was done in
 \ref\ooguri{H. Ooguri and
C. Vafa, {\it All Loop N=2 String Amplitudes},
Nucl. Phys. B451 (1995) 121, hep-th/9505183.} when the four manifold
is $T^2\times R^2$, i.e. when considering Type II strings compactified
on $T^2$.  The amplitudes thus obtained involved Eisenstein functions
of various degrees (as a function of complex/kahler structure of $T^2$).
Unlike the four-dimensional $R^2 F^{2g-2}$ terms, the
$R^4 H^{4g-4}$ (or $R^4 F^{4g-4}$) terms survive in the large volume
limit to give non-zero contributions for the uncompactified superstring.
Furthermore, the $R^4$ term at genus one has precisely the same index
structure as the $R^4$ term multiplying $E_{3/2}(\tau)$ in the conjecture
of \Green.

As will be discussed in this paper, the structure of the $R^4 H^{4g-4}$
terms leads us to conjecture that they are multiplied by the
manifestly SL(2,Z)-invariant Eisenstein function $E_{g+\half} (\tau)$
in the uncompactified Type IIB low-energy effective action.
More precisely, we conjecture that
there is a term in the Type IIB effective action on $R^{10}$ which in the
Einstein gauge takes the form
\eqn\iib{ {\cal S}= N_g \int d^{10} x \sqrt{det_{10} g} }
$$\sum_{p=2-2g}^{2g-2} (-1)^p
R^4 (H^+)^{2g-2+p} (H^-)^{2g-2-p} {\sum_{m,n}}'
{{\tau_2^{g+\half}}\over{(m+n\tau)^{g+\half+p}(m+n\bar\tau)^{g+\half-p}}} $$
where $H^+ = \tau_2^{-\half} (H_{RR} -\tau H_{NS-NS})$,
$H^- = \tau_2^{-\half}(H_{RR} -\bar\tau H_{NS-NS})$
and $N_g$ is an overall normalization constant.

Furthermore, we conjecture that for compactification on $T^2$,
the eight-dimensional $R^4 H^{4g-4}$ terms are multiplied by
a manifestly SL(3,Z)-invariant version of the
Eisenstein function which generalizes the $R^4$ conjecture of Kiritsis
and Pioline\Kiritsis.
The conjectured form of the amplitude in this case
essentially follows from extending $SL(2,Z)$-invariant
Eisenstein functions obtained by summing over a 2d lattice to
$SL(3,Z)$-invariant functions obtained by
summing over 3d lattice points, and is very natural (and perhaps
unique).   Our conjecture in this case implies that $R^4 H^{4g-4}$
terms
only get perturbative contributions at genus 0 and genus $g$,
just as in $R^{10}$.  Moreover, the genus $g$ contribution
is itself an Eisenstein function of the Kahler structure of $T^2$
and precisely coincides with the corresponding
string computation of \ooguri\ for compactification on $T^2$ for all
$g$.  This we consider strong evidence for our conjecture.

The organization of this paper is as follows:
In section 2, we review the results in
\us\ for six-dimensional topological amplitudes $R^4 H^{4g-4}$
(or $R^4 F^{4g-4}$) which arise upon compactification
of Type IIB (or Type IIA) on a hyper-Kahler manifold.  We also discuss the
corresponding computation of $R^2F^{2g-2}$ terms
on Calabi-Yau threefolds, in
part to contrast it with the computations of $R^4 H^{4g-4}$ terms.
 In section 3, we review the topological computations of \ooguri\
when the four manifold is $T^2\times R^2$,
and describe
their implications for scattering amplitudes of Type II
upon compactification on $T^2$ down to $D=8$.
In section 4, we conjecture the non-perturbative
structure of Type IIB $R^4 H^{4g-4}$ terms in $D=8$ and $D=10$, and describe
various types of evidence for our conjecture.
In section 5, we
discuss a paradox concerning the non-perturbative structure
of Type IIA $R^4 F^{4g-4}$ terms and suggest a possible resolution.
In section 6, we discuss
possible implications of these results for ``pair creation'' of
$(p,q)$ strings (motivated by the implications of $R^2F^{2g-2}$
terms for Schwinger's pair creation).
In the concluding section,
we discuss possible implications of our conjecture for the
$\nabla^n R^4$ conjectures of Russo\russo\ and for the
$F^{2g+2}$ conjectures coming
from M(atrix) theory \ref\doug{K. Becker and M. Becker,
{\it A Two-Loop Test of M(atrix) Theory}, Nucl. PHys. B506 (1997)
48, hep-th/9705091\semi M. Douglas, private communication\semi
H. Ooguri, private communcation\semi K. Becker, M. Becker,
J. Polchinski and A. Tseytlin, {\it Higher Order Graviton Scattering
in M(atrix) Theory}, Phys. Rev. D56 (1997) 3174, hep-th/9706072.}.

\newsec{Review of Topological Amplitudes}

In reference \us, we proved that certain six-dimensional superstring
scattering amplitudes can be expressed as topological computations on
the hyper-Kahler compactification manifold.
Although our proof used the modified Green-Schwarz formalism where
spacetime-supersymmetry is manifest and twisting is natural,
it should be straightforward to reproduce our proof using the
Ramond-Neveu-Schwarz formalism. Like their four-dimensional counterparts,
the six-dimensional topological amplitudes involve the scattering of
gravitons and Ramond-Ramond fields, and to understand their structure,
it will be useful to first review the four-dimensional case.

\subsec{Review of four-dimensional $R^2 F^{2g-2}$ terms}

Four-dimensional $R^2 F^{2g-2}$ terms in the effective action
of the Type II
superstring compactified on a Calabi-Yau manifold can be
computed by scattering two gravitons and $2g-2$ chiral graviphotons \topol.
For the Type IIB (or Type IIA) superstring, the vertex operator
for each chiral graviphoton carries $+3/2$ left-moving charge
and $+3/2$ (or $-3/2$) right-moving charge with respect
to the left and right-moving U(1) generators of the N=2 c=9
superconformal field theory representing the compactification.

At $g$ loops, one needs $3g-3$ left and right-moving
picture-changing operators if the
$2g-2$ graviphotons are all chosen in the $(-\half,-\half)$ picture. By U(1)
conservation of the Type IIB (or Type IIA) superstring,
the only contributing part of the picture-changing
operators is $e^{\phi_L} G_L^-$ and $e^{\phi_R} G_R^-$ (or
$e^{\phi_L} G_L^-$ and $e^{\phi_R} G_R^+$), where $\phi_{L/R}$ comes
from fermionizing the $\beta_{L/R}$ ghosts
and $G^{\pm}_{L/R}$ are the fermionic generators of the N=2
c=9 algebra. The spacetime-dependent
part of the computation is trivial, leaving only a correlation function
over $G_L^-$'s and $G_R^-$'s (or
$G_L^-$'s and $G_R^+$'s) which is the N=2 topological computation for
the ``B-model'' (or ``A-model'').

The final result is that the low-energy effective action contains
a local $g$-loop contribution given by the N=2 D=4 superspace expression
\eqn\superfour{\int d^4 x
\int d^2 \theta_L d^2\theta_R (W_{\a\b} W^{\a\b})^g f_g}
where $W_{ab}= (\s^{\mu\nu})_{\a\b} F_{\mu\nu} +(\theta_L \s^{\mu\nu})_\a
(\theta_R \s^{\mu\nu})_\b R_{\mu\nu\rho\kappa}  + ... $,
$F_{\mu \nu}$ denotes the graviphoton field strength,
and $f_g$ is
the topological amplitude at genus $g$ which
depends on the moduli of Calabi-Yau compactification. Integration over
$\theta_{L,R}$ is easily seen to
give $R^2 F^{2g-2}$ terms contracted in various ways, as well as other terms
which are related by supersymmetry.
In four-dimensional Einstein gauge, i.e. ${\cal S}=\int d^4 x \sqrt{det_4 g}
(R + F^2 + ...)$,
it is easy to check there are no $e^\phi$ factors in front of the
$R^2 F^{2g-2}$ term at genus $g$.
This is explained by the fact that the dilaton
in Einstein gauge sits in a tensor multiplet, which cannot appear in
the chiral superspace action of \superfour. Therefore, the $R^2 F^{2g-2}$
term gets no perturbative or non-perturbative contributions except at genus
$g$.\foot{Note that superstring arguments alone can only prove the absence of
contributions below genus $g$ (where there are not enough picture-changing
operators to absorb the U(1) charge), but cannot prove the absence of
contributions above genus $g$.}

\subsec{Topological amplitudes in six dimensions}

As shown in \us, there is a six-dimensional analog of the
four-dimensional amplitudes which involves the genus $g$ scattering
of four
gravitons and $4g-4$ Ramond-Ramond fields. In 6 dimensions,
the Lorentz group is most conveniently described
using the spin group which is $SU(4)$.  We will denote
$SU(4)$ indices by $a,b=1,...,4$, which can describe
either chiral or anti-chiral spinors.  Moreover, spinors carry
an internal $SU(2)$ index denoted by $j,k=\pm$ which comes
from the $SU(2)$ of the hyper-Kahler manifold.\foot{
For $T^4$, the internal
directions will involve an $SO(4)=SU(2)\times SU(2)$. If we
are dealing with $K3$, one of the internal $SU(2)$'s is broken
by the holonomy of $K3$ and only one survives.}
Bispinor
Ramond-Ramond field strengths $M_{ab}^{jk}$ carry both left and
right-moving version of these $SU(4)$ and $SU(2)$ indices.
Note
that for the Type IIB (or Type IIA) superstring,
$b$ has the same (or opposite) chirality as $a$.
So in vector notation, $M_{ab}^{jk}$ describes field-strengths with an
odd (or even) number of vector indices and we will focus primarily on
the three-form $H$ (or two-form $F$).

Now,
suppose that we consider
$4g-4$ Ramond-Ramond
vertex operators and have all of them carry $+1$ left-moving charge and $-1$
right-moving
charge with respect to the left and right-moving U(1) generators of the
N=2 c=6 superconformal field theory representing the
compactification. This implies that the $SU(2)$ indices on the
Ramond-Ramond field strengths $M_{ab}^{jk}$
are all chosen to be in the directions
$j=+$ and $k=+$ where, for later convenience, we choose notation such
that the right-moving SU(2) index on $M_{ab}^{jk}$
has
the opposite sign of the right-moving U(1) charge.

If all Ramond-Ramond vertex operators
are chosen in the $(-\half,-\half)$ picture, one needs $4g-4$
picture-changing operators and by U(1) conservation, only the
$e^{\phi_L} G_L^-$ and $e^{\phi_R} G_R^+$ terms in the picture-changing
operators contribute.
As before, the spacetime-dependent
part of the computation is trivial, leaving only a correlation function
involving $G_L^-$'s and $G_R^+$'s, which is a topological computation on the
hyper-Kahler compactification manifold.

The result is that the $g$-loop six-dimensional low-energy effective action
contains a term proportional to the N=2 D=6
superspace expression
\eqn\supersix{\int d^6 x \int d^4 \theta_L^+ d^4\theta_R^+ (W^{++}_{a_1 b_1}
W^{++}_{a_2 b_2}
W^{++}_{a_3 b_3}
W^{++}_{a_4 b_4} \epsilon^{a_1 a_2 a_3 a_4}
\epsilon^{b_1 b_2 b_3 b_4} )^g ~f_g}
where $W^{++}_{\a\b}=  M^{++}_{ab} +(\theta^+_L \s^{\mu\nu})_a
(\theta^+_R \s^{\mu\nu})_b R_{\mu\nu\rho\kappa}  + ... $ and $f_g$ is
the topological $N=4$ partition function at genus $g$ and instanton number
$(2g-2,2g-2)$, which is
the same as the $N=2$ string partition function at genus $g$ and
instanton number $(2g-2,2g-2)$ on the corresponding
hyper-Kahler manifold.
Note that by U(1) conservation in the $N=2$ current
algebra, this amplitude vanishes when the genus
is less than $g$.

Although the above computation breaks the internal SU(2) invariance to
a U(1) subgroup, the full SU(2) is easily restored by introducing
the ``harmonic'' variables $u_j$ and $\bar u_j$ satisfying
$u_{[j} \bar u_{k]}=\epsilon_{jk}$ \ref\harm{A. Galperin,
E. Ivanov, S. Kalitzyn, V. Ogievetskii and E. Sokatchev,
{\it Unconstrained N=2 Matter, Yang-Mills and Supergravity
Theories in Harmonic Superspace}, Class. Quant. Grav. 1 (1984) 469.}.
Actually, since
we are considering bispinors, we need to introduce two harmonic
variables, $(u^L_j,\bar u^L_j)$ and $(u^R_j,\bar u^R_j)$. By defining
$\hat M_{ab}= u^L_j u^R_k M^{jk}_{ab}$,
we can now repeat the above calculation for arbitrary values of
$u^L_j$ and $u^R_j$, where $\hat M_{ab}$ now includes all the field
strengths which are related to each other by the $SU(2)$ internal rotation.

The spacetime dependence of the computation is still trivial, and the
compactification dependent
topological computation $f_g(\bar u^L_j,\bar u^R_j)$
is now a polynomial of
degree $4g-4$ in $\bar u^L_j$ and $\bar u^R_j$. As shown in \us,
\eqn\inst{f_g(\bar u^L_j, \bar u^R_j)=}
$$ \sum_{n_L,n_R=2-2g}^{2g-2}
F_g^{n_L,n_R}(\bar u^L_+)^{2g-2+n_L}
(\bar u^L_-)^{2g-2-n_L}
(\bar u^R_+)^{2g-2+n_R}
(\bar u^R_-)^{2g-2-n_R}$$
where $F_g^{n_L,n_R}$ computes the $g$-loop partition function of
(left,right) instanton number $(n_L,n_R)$ for the self-dual $N=2$ string
propagating on the hyper-Kahler four manifold.

Knowing the scattering amplitude for any value of $u^L_j$ and $u^R_j$
allows one to construct the SU(2)-invariant amplitude by integrating
over $u^L_j$ and $u^R_j$ as in \harm.
So the complete SU(2)-invariant scattering amplitude is given by the
superspace expression
\eqn\superharm{\int d^6 x \int du^L \int du^R
\int d^4\theta_L d^4\theta_R }
$$(\hat W_{a_1 b_1}
\hat W_{a_2 b_2}
\hat W_{a_3 b_3}
\hat W_{a_4 b_4} \epsilon^{a_1 a_2 a_3 a_4}
\epsilon^{b_1 b_2 b_3 b_4} )^g  f_g( \bar u^L_j, \bar u^R_j) $$
where $\hat W_{\a\b}=  \hat M_{ab} +(\theta_L \s^{\mu\nu})_a
(\theta_R \s^{\mu\nu})_b R_{\mu\nu\rho\kappa}  + ... $,
and $\int du_L\int du_R$ is defined by
\eqn\intha{\int du_L \int du_R ~f^{j_1 ... j_N\,
k_1 ... k_N}
g^{l_1 ... l_N\, m_1 ... m_N}
u_{j_1}^L u_{k_1}^R
\bar u_{l_1}^L \bar u_{m_1}^R ...
u_{j_N}^L u_{k_N}^R \bar u_{l_N}^R \bar u_{m_N}^R }
$$= f^{(j_1 ... j_N)\, (k_1 ... k_N)}
g_{j_1 ... j_N\, k_1 ... k_N}.$$
Expanding \superharm\ in components for the Type IIB (or Type IIA)
superstring gives
$R^4 H^{4g-4}$ terms (or $R^4 F^{4g-4}$ terms), as well as various
other terms related by supersymmetry.

\newsec{Terms in the Eight-Dimensional Effective Action}

Although $f_g(\bar u^L_j,\bar u^R_j)$ is unknown when the
compactification manifold is $K3$, it is known \ooguri\ up to an overall
constant for all $g$ when
the manifold is $R^2\times T^2$, i.e. when the superstring is compactified
to eight dimensions on $T^2$. So by `Lorentz-covariantizing' the
six-dimensional indices of \superharm\ to eight-dimensional indices,
one can find explicit expressions for $g$-loop
terms in the eight-dimensional low-energy effective
action of the Type II superstring.

To `Lorentz-covariantize', one first rewrites \superharm\ in
six-dimensional vector
notation by replacing
$\hat W_{ab}$
with $\sum_n \hat W_{\mu_1 ... \mu_n} \Gamma^{\mu_1 ... \mu_n}_{ab}$
and using
$$\epsilon^{a_1 a_2 a_3 a_4}
\epsilon^{b_1 b_2 b_3 b_4}=\d^{[a_1}_{[b_1}
\d^{a_2}_{b_2}
\d^{a_3}_{b_3}
\d^{a_4]}_{b_4]}$$
 to get
traces of $\Gamma$ matrices.
It is easy to check that the expression only involves contractions
of vector indices and contains no six-dimensional $\epsilon$-tensors.
One now simply replaces all six-component vector indices with eight-component
vector indices.

The only subtlety is that the $u_+^L u_-^R$ and
$u_-^L u_+^R$ pieces of $\hat M_{ab}$
come from
eight-dimensional fields containing indices in the 7 or 8 directions.
For example, in the Type IIB (or Type IIA) superstring
$M_{ab}^{+-} \G_{\mu\nu\rho}^{ab}$
(or $M_{ab}^{+-} \G_{\mu\nu}^{ab}$)
comes from
a four-form (or three-form) in eight dimensions with one component in the
$(7+i8)$ direction. For this reason, we shall
restrict our attention for the rest of this paper to fields coming
from the $u_+^L u_+^R$ and $u_-^L u_-^R$ pieces of $\hat
M_{ab}$, which contain
the same number of indices in six and eight dimensions. For terms
involving these fields, the expression obtained by replacing
six-component with eight-component indices is manifestly Lorentz-invariant
in eight dimensions.

For compactification of Type II strings on $T^2$, the massless bosonic fields
are described by
the Riemann tensor $R_{\mu\nu\rho\sigma}$, a real triplet of three-form
field-strengths
$H_{\mu\nu\rho}^{(jk)}$ where $j,k=\pm$ are SU(2) indices,
a complex triplet of two-form field-strengths
$F_{\mu\nu}^{\pm(jk)}$, a self-dual and anti-self-dual
four-form field-strength $F_{\mu\nu\rho\sigma}^{\pm}$,
and seven scalars consisting of
the $T^2$ Kahler modulus $\s=\s_1 +i \s_2$, the $T^2$ complex modulus
$\rho=\rho_1+i\rho_2$, the eight-dimensional dilaton $\l_8$, and two
Ramond-Ramond scalars.
In terms of these fields,
\eqn\expan{IIB: \hat M_{ab} =
u_+^L u_+^R \G^{\mu\nu\rho}_{ab} H_{\mu\nu\rho}^{(++)} +
u_-^L u_-^R \G^{\mu\nu\rho}_{ab} H_{\mu\nu\rho}^{(--)} + ...,}
$$
IIA: \hat M_{ab} =
u_+^L u_+^R \G^{\mu\nu}_{ab} F_{\mu\nu}^{+(+-)} +
u_-^L u_-^R \G^{\mu\nu}_{ab} F_{\mu\nu}^{-(+-)} + ... , $$
where the terms in $...$ will be ignored.
Note that $H_{\mu\nu\rho}^{(+-)}$,  $F_{\mu\nu}^{\pm(++)}$,
and $F_{\mu\nu}^{\pm(--)}$ are NS-NS fields which do not appear
in $\hat M_{ab}$.

In eight-dimensional Einstein gauge, i.e.
$${\cal S}=\int d^8 x \sqrt{det_8 g}~ (R +
F_{\mu\nu \, +}^{(jk)}
F^{\mu\nu}_{-(jk)} +
H_{\mu\nu\rho}^{(jk)}
H^{\mu\nu\rho}_{(jk)} + ...),$$
the topological computation of \ooguri\ found
\eqn\dependence{f_g(\bar u_j^L,\bar u_j^R)
= \l_8^{{{2g-2}\over 3}} \s_2^g F_g }
where
$\l_8 = \s_2^{-\half} e^\phi$ is the eight-dimensional coupling constant,
$\s_2$ is the volume of $T^2$,
and (up to an overall constant)
\eqn\comp{F_g (\bar u^L_j, \bar u^R_j) = {\sum_{m,n}}' ( {{\bar u_+^L
\bar u_+^R}\over
{m + n \sigma}}
+ {{\bar u_-^L \bar u_-^R}\over
{m + n \bar\sigma}} )^{4g-4} | m + n\sigma|^{2g-4}}
where $\sum'$ means to sum over all integers $m$ and $n$ except when
$m=n=0$.
The $e^{{2\over 3} (g-1)\phi}$ dependence of $f_g$ can be understood
by rescaling
\eqn\rescal{g_{\mu\nu} \to e^{-2\phi/3} g_{\mu\nu},\quad
F_{\mu\nu}^{\pm(jk)}
\to e^{-\phi/3} F_{\mu\nu}^{\pm(jk)}, \quad
H_{\mu\nu\rho}^{(jk)} \to e^{-2\phi/3}
H_{\mu\nu\rho}^{(jk)} }
which rescales the eight-dimensional Einstein gauge
action to string gauge and rescales
\eqn\newres
{ e^{{2\over 3} (g-1)\phi} \sqrt{det_8 g}~ R^4 (H^{4g-4} + F^{4g-4}) \to
e^{2(g-1)\phi} \sqrt{det_8 g}~ R^4 (H^{4g-4} + F^{4g-4}) .}

The topological computation of \comp\
was called the `A-model' in \ooguri.
For
compactification on $T^2\times R^2$, the N=2 U(1) generator splits as
$J= J_1 + J_2$ where $J_1$ comes from $T^2$ and $J_2$ comes from $R^2$.
The topological computation for the `B-model' comes from flipping the
sign of the right-moving $J_1$ with respect to the right-moving $J_2$.
This flip is just a $T$-duality transformation on one of the circles in
$T^2$, so the `B-model' computation is related to the `A-model' computation
by replacing $\s$ with $\rho$ in \comp\ and $\s_2$ with $\rho_2$ in
\dependence. Since the `B-model' computation vanishes as $\s_2\to\infty$, it
does not survive in ten dimensions. In eight dimensions, the scattering
amplitudes associated with the
`B-model' involve $R^4 F^{4g-4}$ (or $R^4 H^{4g-4}$)
terms in the Type IIB (or Type IIA) superstring
effective action.

\subsec{$SL(2,Z)\times SL(2,Z)$ invariance}

For compactification on $T^2$, the perturbative
low-energy effective action is
invariant under $SL(2,Z)\times SL(2,Z)$ transformations.
As usual, it is useful to think of the Kahler and complex moduli
as $SL(2,R)/SO(2)$ variables, $c_I^j$ and $\tilde c_I^j$ ($j=\pm$ and
$I=1$ or 2). These variables satisfy
\eqn\ms{{\cal M}_{IJ}
\equiv c_{(I}^+ c_{J)}^- = {1\over \s_2}\left(\matrix{1&\s_1\cr
\s_1&|\s|^2\cr}\right) , }
\eqn\mstwo{\tilde {\cal M}_{IJ}\equiv \tilde c_{(I}^+ \tilde c_{J)}^-
= {1\over \rho_2}\left(\matrix{1&\rho_1\cr
\rho_1&|\rho|^2\cr}\right) ,}
where ${\cal M}_{IJ}\to \Lambda_I^K \Lambda_J^L {\cal M}_{KL}$ and
$\tilde {\cal M}_{IJ}\to \tilde\Lambda_I^K \tilde\Lambda_J^L
\tilde{\cal M}_{KL}$ under
$SL(2,Z)\times SL(2,Z)$
transformations parameterized by $\Lambda_I^J$ and $\tilde\Lambda_I^J$.
Here $\sigma$ denotes the Kahler class of $T^2$ and $\rho$ denotes
its complex structure.

Under $SO(2)\times SO(2)$ transformations,
$c_I^\pm \to e^{\pm i\theta} c_I^\pm$
and $\tilde c_I^\pm \to e^{\pm i\tilde\theta} \tilde c_I^\pm$, so one
can choose an SO(2) gauge in which
\eqn\gaugech{c_I^+ =\s_2^{-1/2} (1,\s), \quad
c_I^- =\s_2^{-1/2} (1,\bar\s), \quad
\tilde c_I^+ =\rho_2^{-1/2} (1,\rho), \quad
\tilde c_I^- =\rho_2^{-1/2} (1,\bar\rho).}
The relevant Ramond-Ramond
field strengths are defined to be invariant under
$SL(2,Z)\times SL(2,Z)$ transformations, however they transform under
$SO(2)\times SO(2)$ transformations for the Type IIB (or Type IIA)
superstring as
\eqn\tran{H_{\mu\nu\rho}^{(\pm\pm)} \to e^{\pm i\theta}
H_{\mu\nu\rho}^{(\pm\pm)}, \quad
F_{\mu\nu}^{\pm(+-)} \to e^{\pm i\tilde \theta}
F_{\mu\nu}^{\pm(+-)} }
$$(or ~
H_{\mu\nu\rho}^{(\pm\pm)} \to e^{\pm i\tilde\theta}
H_{\mu\nu\rho}^{(\pm\pm)}, \quad
F_{\mu\nu}^{\pm(+-)} \to e^{\pm i\theta}
F_{\mu\nu}^{\pm(+-)} ~).$$

The eight-dimensional Einstein gauge
action obtained by `covariantizing' \superharm\ is
invariant under these transformations since it can be written as
\eqn\eighta{{\cal S}=\int d^8 x \int du^L du^R \sqrt{det_8
g}~ \l_8^{{2g-2}\over 3}
R^4 \hat M^{4g-4}  }
$$
{\sum_{m^1, m^2}}' ( {{\bar u_+^L \bar u_+^R}\over
{m^I c_I^+}}
+ {{\bar u_-^L \bar u_-^R}\over
{m^I c_I^-}} )^{4g-4} | m^I c_I^+ |^{2g-4}$$
where $\hat M_{ab}$ is defined as in \expan\ and the
index contractions on
$R^4 \hat M^{4g-4}$ are determined using the method discussed earlier.
Invariance is manifest if
$m^I\to
(\Lambda^{-1})^J_I m^I$ under
$SL(2,Z)\times SL(2,Z)$ transformations
and $(u_\pm^L,
u_\pm^R)$
 $\to $
$(e^{\pm{i\over 2}(
\theta +\tilde\theta)} u_\pm^L,
e^{\pm{i\over 2}(
\theta -\tilde\theta)} u_\pm^R)$
under
$SO(2)\times SO(2)$ transformations. Similar techniques 
have previously been used in \ref\noholom{A. Kehagias and
H. Partouche, {\it On the Exact Quartic Effective Action for the
Type IIB Superstring}, hep-th/9710023\semi 
M.B. Green, M. Gutperle and H. Kwon, {\it Sixteen-Fermion and
Related Terms in M-Theory on $T^2$,} hep-th/9710151\semi
A. Kehagias and
H. Partouche, {\it D-Instanton Corrections as (p,q)-String
Effects and Non-Renormalization Theorems}, hep-th/9712164.}.  

Performing the integrations over the $u^L$ and $u^R$ variables, one
obtains for the Type IIB superstring
\eqn\eigb{{\cal S}=\int d^8 x \sqrt{det_8 g}~ \l_8^{{2g-2}\over 3}
\sum_{p=2-2g}^{2g-2} R^4 (H^{(++)})^{2g-2+p} (H^{(--)})^{2g-2-p}  }
$$
{\sum_{m, n}}' {\s_2^g \over{(m + n\sigma)^{g+p}
(m + n\bar\sigma)^{g-p}}} .$$
Note that in terms of the D=10
three-form and five-form field strengths,
comparison of the D=8 and D=10
kinetic terms implies
that
\eqn\deften{H^{(\pm\pm)}_{\mu\nu\rho}=
\s_2^{1/3} \tau_2^{-1/3} (H^{R-R}_{\mu\nu\rho} -\tau_1 H_{\mu\nu\rho}^{NS-NS})
\mp
i\s_2^{4/3}\tau_2^{2/3} H^{R-R}_{\mu\nu\rho\, 8 9} .}
For the Type IIA superstring, the only difference
from \eigb\ is that $~~~~~~~~~~$
$(H^{(++)})^{2g-2+p}
(H^{(--)})^{2g-2-p}  $ is replaced with $~~~~~~~~~~$
$(F^{+(+-)})^{2g-2+p}$
$(F^{-(+-)})^{2g-2-p} $ where, in terms of the
D=10 Ramond-Ramond two-form and four-form field strengths,
$$F^{\pm(+-)}_{\mu\nu}=\s_2^{1/6} \tau_2^{-2/3} F^{R-R}_{\mu\nu} \pm
i\s_2^{7/6}\tau_2^{1/3} F^{R-R}_{\mu\nu\, 8 9} .$$

\newsec{Type IIB $R^4 H^{4g-4}$ Conjectures}

\subsec{Eight-dimensional conjecture}

For the Type IIB (or Type IIA)
superstring compactified on $T^2$, it has been conjectured that
the $SL(2,Z)\times SL(2,Z)$ symmetry of the previous subsection
is extended non-perturbatively to an $SL(3,Z)\times SL(2,Z)$
symmetry where the $SL(3,Z)$ extends the Kahler (or complex) moduli
of $T^2$.
In this subsection, we conjecture an
$SL(3,Z)\times SL(2,Z)$-invariant
non-perturbative extension of \eighta\ for the case of $R^4 H^{4n-4}$
terms in the Type IIB superstring effective action.
When $n=1$, our conjecture coincides
with that of \Green\ and \Kiritsis, and when $n>1$, it generalizes
their
Type IIB conjecture in a natural way.

Under
$SL(3,Z)\times SL(2,Z)$ transformations, the seven scalars of the
Type IIB superstring on $T^2$ split into
a quintuplet which can be thought of as $SL(3,R)/SU(2)$ variables
$C_\alpha^{(jk)}$ ($\alpha=1$ to 3 and $j,k=\pm$), and
a doublet which can be thought of as $SL(2,R)/SO(2)$ variables
$\tilde c_I^j$. $\tilde c_I^j$ is defined as in \gaugech\ and \mstwo.
$C_\a^{(jk)}$ is defined to satisfy \Kiritsis:
\eqn\matrt{{\cal M}_{\a\b}\equiv
C_\a^{(jk)} C_{\b\,(jk)} =}
$$= \l_8^{4/3} \left(\matrix{1 & \tau_1
&B_R + \tau_1 \s_1 \cr \tau_1
& |\tau|^2
& \tau_1 B_R + |\tau|^2 \s_1  \cr  B_R + \tau_1 \s_1
& \tau_1 B_R + |\tau|^2 \s_1 &
(B_R +\tau_1 \s_1)^2 + \l_8^{-2} \s_2^{-1} |\s|^2\cr }\right) $$
where $\tau=\tau_1 +i \l_8^{-1} \s_2^{-\half}$,
$\tau_1$ and $B_R$ are the two Ramond-Ramond scalars ($\tau_1$ is already
a scalar in 10 dimensions and $B_R$ is the constant expectation value
of the RR B-field on $T^2$), and
${\cal M}_{\a\b}\to \Lambda_\a^\g \Lambda_\b^\d{\cal M}_{\g\d}$
under $SL(3,Z)$
transformations parameterized by $\Lambda_\a^\b$.

Under $SU(2)$ transformations,
$C_\a^{(jk)} \to \Omega^j_l \Omega^k_m C_\a^{(lm)}$,
so one can choose an SU(2) gauge in which
\eqn\gaugecht{C_\a^{(++)} =\l_8^{-1/3}\s_2^{-1/2} (0,1,\s ), \quad
C_\a^{(--)} =\l_8^{-1/3}\s_2^{-1/2} (0,1,\bar\s), }
$$C_\a^{(+-)} = -i\l_8^{2/3} (1,\tau_1, B_R +\tau_1\s_1).$$
The triplet of three-form
field strengths are defined to be invariant under
$SL(3,Z)\times SL(2,Z)$ transformations, however they transform under
$SU(2)\times SO(2)$ transformations as
$H_{\mu\nu\rho}^{(jk)} \to \Omega^j_l \Omega^k_m
H_{\mu\nu\rho}^{(lm)}$.

For the Type IIB superstring, the Ramond-Ramond $H_{\mu\nu\rho}$
fields appear in $\hat M_{ab}$ as in \expan. To make SU(2) invariance
manifest, it is useful to define a new harmonic field
\eqn\newh{\hat{\widetilde M}_{ab}(u_j^L, u_j^R)
 = \G^{\mu\nu\rho}_{ab} H_{\mu\nu\rho}^{(jk)} u_j^L
u_k^R}
which now contains the NS-NS three-form field strength.
Note that $H^{(+-)}_{\mu\nu\rho}$ is imaginary since the
reality condition is $(H^{(jk)}_{\mu\nu\rho})^*$
=$\epsilon_{jl}\epsilon_{km}H^{(jk)}_{\mu\nu\rho}$. 
In terms of the D=10
three-form field strengths,
$H^{(+-)}_{\mu\nu\rho}=i \s_2^{1/3} \tau_2^{2/3} H^{NS-NS}_{\mu\nu\rho}$, and
$H^{(\pm\pm)}_{\mu\nu\rho}$ are
defined as in \deften.

Our $SL(3,Z)\times SL(2,Z)$-invariant
conjecture for $R^4 H^{4g-4}$ terms in the eight-dimensional
Einstein gauge Type IIB low-energy effective action is
\eqn\conjb{{\cal S} = N_g
\int d^8 x \int du^L \int du^R \sqrt{det_8 g}~ R^4
\hat {\widetilde M}^{4g-4}}
$${\sum_{m^1, m^2, m^3}}' ( m^\a C_\a^{(jk)}\bar u_j^L \bar u_k^R )^{4g-4}
(m^\a C_\a^{(jk)} C_{\b (jk)} m^\b)^{-3g +{3\over 2}} $$
where $N_g$ is an overall constant
and $\sum'$ means to sum over all integers $m^1$, $m^2$ and $m^3$ except
when $m_1=m_2=m_3=0$.
Note that when $g=1$, this conjecture coincides with the
D=8 $R^4$ conjecture of \Kiritsis.   Before checking how
this conjecture fits with the perturbative computation in
eight dimensions given in \eigb, let us discuss what
this conjecture implies for the uncompactified Type IIB
effective action in 10 dimensions.

\subsec{Ten-dimensional conjecture}

To obtain our conjecture for ten-dimensional $R^4 H^{4g-4}$ terms,
one takes the large volume limit $\s_2\to\infty$, keeping
$\tau_2 \equiv \l_8^{-1} \s_2^{-\half}$ fixed.
In this large volume limit,
the terms in \conjb\ involving $m^3\neq 0$ vanish, giving
\eqn\contf{{\cal S} =N_g \int d^8 x \int du^L \int du^R \sqrt{det_8 g}
{}~(\s_2^2\tau_2)^{(2g+1)/6}~R^4
\hat {\widetilde M}^{4g-4}}
$${\sum_{m^1, m^2}}' ( m^I D_I^{(jk)}\bar u_j^L \bar u_k^R )^{4g-4}
(m^I D_I^{(jk)} D_{J(jk)} m^J)^{-3g +{3\over 2}} $$
where $I=1$ to 2 and
\eqn\brt{D_I^{(++)}=D_I^{(--)}= \tau_2^{-\half} (0,\tau_2),\quad
D_I^{(+-)}= -i\tau_2^{-\half} (1,\tau_1).}

It is convenient to define $H^\pm_{\mu\nu\rho}=$
$\s_2^{-1/3}\tau_2^{-1/6} [$
$ \half (H_{\mu\nu\rho}^{(++)} +
H_{\mu\nu\rho}^{(--)}) \pm$
 $
H_{\mu\nu\rho}^{(+-)} ]$, which can be expressed in terms of the $D=10$
NS-NS and R-R three-forms as
$H^+_{\mu\nu\rho}=\tau_2^{-\half} (H^{R-R}_{\mu\nu\rho} -
\tau H^{NS-NS }_{\mu\nu\rho} )$ and
$H^-_{\mu\nu\rho}=\tau_2^{-\half} (H^{R-R}_{\mu\nu\rho} -
\bar\tau H^{NS-NS }_{\mu\nu\rho}) $.
Rescaling to ten-dimensional Einstein gauge (where the
classical action after compactification on $T^2$ is
${\cal S}= N_g \int d^8 x \sqrt{det_8 g}~ \s_2 \tau_2^{\half}
{}~(R + H_{\mu\nu\rho}^+ H^{-\mu\nu\rho} + ...$), one obtains
\eqn\contg{{\cal S} =N_g\int d^8 x \int du^L \int du^R \sqrt{det_8 g}
{}~\s_2\tau_2^{\half} R^4
(\hat {\widetilde M}^{4g-4} \s_2^{-1/3} \tau_2^{-1/6} )^{4g-4}}
$${\sum_{m^1, m^2}}' ( m^I D_I^+ \bar v_+^L \bar v_+^R
-
m^I D_I^- \bar v_-^L \bar v_-^R )^{4g-4}
(m^I D_I^+ D_J^- m^J)^{-3g +{3\over 2}} $$
where $D_I^+ =\tau^{-\half}(1,\tau)$,
$D_I^- =\tau^{-\half}(1, \bar\tau)$, $\bar v_\pm^L =
2^{-\half}(\bar u_+^L \pm \bar u_-^L)$,
and $\bar v_\pm^R = 2^{-\half}(\bar u_+^R \pm \bar u_-^R)$ .

Ignoring
$H_{\mu\nu\rho}^{(++)} -
H_{\mu\nu\rho}^{(--)}$ (which comes from the dimensional reduction of
the five-form), converting all contracted eight-component vector indices
into contracted ten-component vector indices, and integrating over
$u$, one obtains
our manifestly $SL(2,Z)$-invariant
ten-dimensional conjecture for the Type IIB $R^4 H^{4g-4}$ term
\eqn\iif{ {\cal S}= N_g \int d^{10} x \sqrt{det_{10} g} }
$$\sum_{p=2-2g}^{2g-2} (-1)^p
R^4 {(H^+)}^{2g-2+p} {(H^-)}^{2g-2-p} {\sum_{m,n}}'
{{\tau_2^{g+\half}}\over{(m+n\tau)^{g+\half+p}(m+n\bar\tau)^{g+\half-p}}} .$$

For the term with $p=0$ in \iif,
the coefficient multiplying $R^4 H^4$
is proportional to the
Eisenstein function
\eqn\eiso{E_{g+\half}(\tau)\equiv{1\over{2\zeta(2g+1)}}{\sum_{m,n}}'
{{\tau_2^{g+\half}}\over{(m+n\tau)^{g+\half}(m+n\bar\tau)^{g+\half}}} .}
Furthermore, the coefficients for $p>0$ are
proportional to 
$\tau_2^{-p}
(\tau_2^2 {\partial\over{\partial\tau}})^p E_{g+\half}(\tau)$
while the coefficients for $p<0$ are proportional to 
$\tau_2^p
(\tau_2^2 {\partial\over{\partial\bar\tau}})^{-p} E_{g+\half}(\tau).$
At large values of $\tau_2$, 
\eqn\eisen{E_{g+\half}(\tau)\to
\tau_2^{g+\half} + \gamma_{g+\half} 
\tau_2^{\half -g} + O (e^{-2\pi\tau_2}) ,}
where $\gamma_{g+\half}$ =
$\sqrt{\pi}{{\G (g) \zeta (2g)}\over{\G (g+\half)\zeta(2g+1)}}$. 
After rescaling
to D=10 string gauge, 
\eqn\rescdt{(\tau_2^{g+\half}+ \gamma_{g+\half} \tau_2^{\half-g})
 \sqrt{det_{10} g} ~R^4 H^4 \to 
(\tau_2^2+ \gamma_{g+\half} \tau_2^{2-2g})
 \sqrt{det_{10} g} ~R^4 H^4 ,}
so \iif\ only gets perturbative contributions at tree-level and at genus $g$. 

\subsec{Evidence for Type IIB conjecture}

The most important evidence for our conjecture comes from explicit
agreement of the eight-dimensional conjecture
with the genus $g$ computation in \eighta.
To compare \conjb\ with \eighta, it is useful to split the sum
${\sum_{m^1, m^2, m^3}}'$ in \conjb\
into $\sum_{m^1}$ where $m_2=m_3=0$, and
$\sum_{m^1} {\sum_{m^2,m^3}}'$ where one sums over all values except
$m_2=m_3=0$. The first sum contributes 
\eqn\conjtr{{\cal S} =N_g\int d^8 x \int du^L \int du^R \sqrt{det_8 g}~ R^4
\hat {\widetilde M}^{4g-4}}
$$\sum_{m^1} ( m^1 C_1^{(+-)}\bar u_{(+}^L \bar u_{-)}^R )^{4g-4}
(- m^1 C_1^{(+-)} C_1^{(+-)} m^1)^{-3g +{3\over 2}} $$
$$ =\int d^8 x \int du^L \int du^R \sqrt{det_8 g}~  \l_8^{-(4g+2)/3} R^4
\hat {\widetilde M}^{4g-4}
(\bar u_{(+}^L \bar u_{-)}^R )^{4g-4}
\sum_{m^1} (m^1)^{-2g-1} $$
where we used that $C_1^{(++)}=C_1^{(--)}=0.$
Integrating over $u$ gives the contribution
\eqn\conjsr{{\cal S} =N_g \int d^8 x \sqrt{det_8 g}~\l_8^{-(4g+2)/3}~ 
\sum_{m^1} (m^1)^{-2g-1} }
$$R^4 ~
\sum_{q=0}^{2g-2} c_q (H^{(++)} H^{(--)})^{2g-2-q} (H^{(+-)})^{2q} $$
where $c_q$ are constants which can be easily computed.
This is easily seen to be a tree-level contribution since the
$\l_8^{-(4g+2)/3}$ dependence 
differs from the $g$-loop $\l_8^{(2g-2)/3}$ dependence of \dependence\
by a factor of $\l_8^{-2g}$. 

To evaluate the contribution of the second sum
$\sum_{m^1} {\sum_{m^2,m^3}}'$, it is convenient to perform a
Poisson resummation on $m_1$. Following \Kiritsis, this can be done by writing
$$\sum_{m^1}{\sum_{ m^2, m^3}}' ( m^\a C_\a^{(jk)}
\bar u_j^L \bar u_k^R )^{4g-4}
(m^\a C_\a^{(jk)} C_{\b (jk)} m^\b)^{-3g +{3\over 2}} $$
$$={\pi^{3g-{3\over 2}}\over
{\Gamma(3g-{3\over 2})}}\int_0^\infty {dt\over t^{3g-{1\over 2}}}
\sum_{m^1}$$
$$
{\sum_{ m^2, m^3}}' ( m^\a C_\a^{(jk)}\bar u_j^L \bar u_k^R )^{4g-4}
\exp (-{\pi\over t} m^\a C_\a^{(jk)} C_{\b (jk)} m^\b)$$
$$={\pi^{3g-{3\over 2}}
\over {\Gamma(3g-{3\over 2})}}\int_0^\infty {dt\over t^{3g-{1\over 2}}}
\sum_{n}\int_{-\infty}^{\infty} dm_1  e^{2\pi i m_1 n}$$
$$ {\sum_{ m^2, m^3}}'
( m^\a C_\a^{(jk)}\bar u_j^L \bar u_k^R )^{4g-4}
\exp (-{\pi\over t} m^\a C_\a^{(jk)} C_{\b (jk)} m^\b)$$
$$={\pi^{3g-{3\over 2}}
\over {\Gamma(3g-{3\over 2})}}\int_0^\infty {dt\over t^{3g-{1\over 2}}}
\sum_{n}\int_{-\infty}^{\infty} dm_1  {\sum_{ m^2, m^3}}' $$
$$
( {1\over{2\pi i}}{d\over dn} C_1^{(jk)}\bar u_j^L \bar u_k^R
+ m^Y C_Y^{(jk)}\bar u_j^L \bar u_k^R )^{4g-4}
\exp (-{\pi \over t} m^\a C_\a^{(jk)} C_{\b (jk)} m^\b +2\pi i m_1 n)$$
\eqn\pois{={\pi^{3g-2}\l_8^{-2/3}\over
{\Gamma(3g-{3\over 2})}}\int_0^\infty {dt\over t^{3g-1}}
\sum_{n} {\sum_{ m^2, m^3}}' }
$$
({1\over{2\pi i}} {d\over dn} C_1^{(jk)}\bar u_j^L \bar u_k^R
+ m^Y C_Y^{(jk)}\bar u_j^L \bar u_k^R )^{4g-4} $$
$$
\exp (-{\pi\over t} m^Y C_Y^{(jk)} C_{Z (jk)} m^Z
-\pi t \l_8^{-4/3} n^2 - 2\pi i n (m_2 \tau_1 +m_3 (B_R +\tau_1\s_1)) )$$
where $Y,Z= 2$ or 3.

Splitting \pois\ into the $n=0$ and $n\neq 0$ parts, it is
straightforward to show that when $n\neq 0$, the contribution
to ${\cal S}$ is of order $O (e^{-\tau_2})$ 
and is therefore non-perturbative.
The contribution to ${\cal S}$ when $n=0$ is proportional to
\eqn\conjc{{\cal S} =\int d^8 x \int du^L \int du^R \sqrt{det_8 g}~
\l_8^{-2/3}~R^4
\hat {\widetilde M}^{4g-4}}
$${\sum_{m^2, m^3}}' ( m^Y C_Y^{(++)}\bar u_+^L \bar u_+^R
+ m^Y C_Y^{(--)} \bar u_-^L \bar u_-^R )^{4g-4}
(m^Y C_Y^{(++)} C_Z^{(--)} m^Z)^{-3g +2}. $$
Comparing $C_Y^{(\pm\pm)}$ with $c_I^\pm$, it is easy to check that
\conjc\ agrees with \eighta\ if one ignores the
NS-NS three-form $H^{(+-)}_{\mu\nu\rho}$ and keeps the R-R
three-forms.
So the conjecture precisely reproduces the $g$-loop computation
of \eighta\ and the only other perturbative contribution to
$R^4 H^{4g-4}$ terms is at tree-level.

There are at least two arguments why a perturbative
non-renormalization theorem
for $R^4 H^{4g-4}$ terms would not be surprising.
One argument comes from superstring amplitude computations which
imply by U(1) conservation that $R^4 (H_{\mu\nu\rho}^{(++)})^{4g-4}$
terms cannot recieve corrections below genus $g$. This does not
disagree with \conjb\ since \conjb\ predicts tree-level
contributions only if there are an equal number of $H_{\mu\nu\rho}^{(++)}$
and
$H_{\mu\nu\rho}^{(--)}$ fields.

The second argument for a non-renormalization theorem comes from the
structure of the duality group and the perturbative decoupling
of Ramond-Ramond zero modes. Since the duality group is
$SL(3,Z)\times SL(2,Z)$, it is reasonable to assume that any
duality-invariant amplitude is proportional to the
factorized product $f(T) g(U)$ where $T$ are the
$SL(2,R)/SO(2)$ moduli and $U$ are the $SL(3,R)/SU(2)$
moduli.\foot{ Of course, the amplitude could be proportional to a sum
of such products $\sum_i f_i(T) g_i(U)$, but in this case, each
term in the sum would have to be separately invariant under
$SL(3,Z)\times SL(2,Z)$ transformations.}
Since the $g$-loop Type IIB $R^4 H^{4g-4}$ amplitude is independent
of the $T$ moduli, $f(T)=1$.
So the full non-perturbative amplitude only depends on the $U$
moduli, which include the two Ramond-Ramond scalars and the string
coupling constant. Although only proven for $R^4$ terms \nonren, it seems
probable that any SL(3,Z)-invariant expression which is perturbatively
independent of the Ramond-Ramond moduli contains only a finite number
of perturbative contributions.

\newsec{ Paradox for Type IIA $R^4 F^{4g-4}$ Term}

In the $g$-loop topological computation,
Type IIA $R^4 F^{4g-4}$ terms
have precisely the same structure as Type IIB $R^4 H^{4g-4}$ terms.
Nevertheless,
we have been unable to find
a natural
$SL(3,Z)\times SL(2,Z)$-invariant extension
of \eighta\ except when $g=1$. When $g=1$, \eighta\ only depends
on the Kahler moduli and is independent of $\l_8$ and the complex
moduli. Therefore, the genus 1 expression in \eighta\ is already
$SL(3,Z)\times SL(2,Z)$-invariant and needs no modification.
But for $g>1$, \eighta\ depends on $\l_8$ which mixes with the
complex moduli under $SL(3,Z)$ transformations.

Note that N=2 D=8 supersymmetry only implies
decoupling of $T$ and $U$ moduli for terms involving eight derivatives
(such as $R^4$ terms) but does not imply decoupling for terms with twelve
derivatives or more.  This is because eight-derivative
superspace actions \nonren\ must be of
the form $\int d^8 x (D_+)^8 (D_-)^8 f(W)$ or
$\int d^8 x \int du (D_+)^8 (\bar D_+)^8 g(L_{++++})$
where $D_\pm$ and $\bar D_\pm$ are
N=2 D=8 supersymmetric derivatives, $W$
is a chiral superfield whose lowest component
is the $T$ modulus, and $L_{jklm}$
is a linear superfield whose lowest components
are the $U$ moduli. But twelve-derivative
superspace actions can be of the form
$\int d^8 x \int du (D_+)^8 (D_-)^8 (\bar D_+)^8 f(W) g(L_{++++})$.

So using the notation
of the previous subsection, the Type IIA
$R^4 F^{4g-4}$ term is multiplied by $f(T) g(U)$ where
$f(T)$ is the $SL(2,Z)$-invariant function
given by the $g$-loop topological computation
and
\eqn\ag{ g(U)=\l_8^{(2g-2)/3}(1 + h(\l_8,\rho, \tau_1, B_R))}
is some $SL(3,Z)$-invariant function.
In order that the $R^4 F^{4g-4}$ term does not blow up in the
ten-dimensional limit, $h(\l_8,\rho,\tau_1, B_R)$ must
go to zero as $\l_8 \to 0$. This implies that the eight-dimensional
$R^4 F^{4g-4}$ term gets no corrections below genus $g$, as expected
from U(1) conservation in the superstring computation.

By taking the $\s_2 \to\infty$ limit where
$\l_8 = \s_2^{-\half} e^{\phi}$, one finds that $h$ does not contribute
so the complete ten-dimensional
$R^4 F^{4g-4}$ term is given by
\eqn\tena{{\cal S}=N_g\int d^8 x \int du^L du^R \sqrt{det_8 g} ~
\s_2^{(2g+1)/3} ~e^{{(2g-2)\phi}\over 3}
R^4 \hat M^{4g-4}  }
$$
(\bar u_+^L \bar u_+^R +\bar u_-^L\bar u_-^R)^{4g-4}
\sum_{m^1\neq 0} (m^1)^{-2g} $$
where
$\hat M_{ab}=
(u_+^L u_-^R +u_-^L u_+^R)\G^{\mu\nu}_{ab} F_{\mu\nu}$ and we are
ignoring
the Ramond-Ramond two-form coming from dimensional reduction of the
D=10 four-form. Replacing all contracted eight-component vector indices
with contracted ten-component vector indices and rescaling to ten-dimensional
string gauge, one obtains the effective action
\eqn\tena{{\cal S}=N_g \int d^{10} x \int du^L du^R \sqrt{det_{10} g}}
$$
e^{(6g-6)\phi}
R^4 \hat M^{4g-4}
(\bar u_+^L \bar u_+^R +\bar u_-^L\bar u_-^R)^{4g-4}
\sum_{m^1\neq 0} (m^1)^{-2g} $$
$$=N_g \int d^{10} x  \sqrt{det_{10} g}~
e^{(6g-6)\phi}
R^4 F^{4g-4}
\sum_{m^1\neq 0} (m^1)^{-2g} $$
up to an overall normalization factor.

If the M-theory conjecture is correct, this term should come from
dimensional reduction of an eleven-dimensional term compactified on
a circle
of radius
$r=e^{2\phi/3}$
where the gauge field $A_\mu$ is identified with $g_{\mu~10}/g_{10~10}$.
As one scales $r$,
\eqn\scale{ \sqrt{det_{10} g} \to r^5 \sqrt{det_{10} g},
\quad R^4 \to r^{-4} R^4,\quad
\hat M^{4g-4} \to r^{8-8g},}
so \tena\ scales like $r^g$, i.e. it blows up faster than the
circle radius when $g>1$
and therefore naively violates the conjecture.

One possible resolution of this paradox
is that the Type IIA $R^4 F^{4g-4}$ term
comes from a non-local term in the eleven-dimensional action,
similar to momentum-dependent Type IIA $R^4$ terms.
As discussed in \tseytlin, the one-loop four-graviton scattering
amplitude in eleven-dimensional supergravity gets contributions
from a local $R^4$ term and from a non-local
$s^{3/2} R^4$ term where $s=p_1\cdot p_2$ is a Mandelstam variable.
After compactification on a circle of radius $r$,
the non-local term gives rise
to an infinite sum of local Type IIA terms
$\sum_{k=2}^\infty c_k r^{2k-2} s^k R^4$,
each of which blows up faster than $r$.
Perhaps the genus $g$ Type IIA $R^4 F^{4g-4}$ term is
the first term of an infinite sum of terms,
$\sum_{k=0}^\infty c_k r^{g+2k} s^k R^4 F^{4g-4}$,
which sums up to a non-local eleven-dimensional term in
the limit $r\to\infty$.\foot{We would like to thank Michael Green
for suggesting that the paradox might be resolved by summing an
infinite series of terms.}
Note that the term proportional to $s^k$ would come from a genus $g+k$
Type IIA term.

\newsec{String ``pair creation'':  A Stringy extension of Schwinger's
Computation}

As we have discussed in section 2,
there are some parallels between the superpotential
$R^2 F^{2g-2}$ terms obtained in the context of Calabi-Yau
threefold compactifications and the $R^4 H^{4g-4}$ terms on which we have
concentrated in this paper.  In the context of Calabi-Yau threefolds,
the $f_g R^2F^{2g-2}$ terms were used in \ref\va{C. Vafa, {\it
A Stringy Test of the Fate of the Conifold}, Nucl. Phys. B447 (1995) 252,
hep-th/9505023.}
\ref\ghosv{D. Ghoshal, C. Vafa, {\it c=1 String as the Topological Theory of
the Conifold}, Nucl. Phys. B453 (1995) 121.}\ref\nareta{I. Antoniadis, E.Gava,
K.S. Narain, T.R. Taylor,
{\it N=2 Type II- Heterotic duality and Higher derivative
F-terms}, Nucl. Phys. B455 (1995) 109.}\ (see
\ref\gopv{R. Gopakumar and C. Vafa, {\it Topological Gravity as
Large N Topological Gauge Theory}, hep-th/9802016.}\ for a recent discussion)
to check Strominger's conjecture
about the resolution of the conifold singularity by a light wrapped
D3 brane \ref\stromin{A. Strominger, {\it Massless Black Holes and Conifolds in
String
Theory}, Nucl. Phys. B451 (1995) 96.}.
In particular, if one considers giving a vev to $F=\lambda $ and computes
the $R^2$ term in the four-dimensional effective action, one gets
\eqn\vevp{S=\int d^4 x \sqrt {det_4 g}~ R^2 f(\lambda) }
where
$$f(\lambda )=\sum_g f_g \lambda^{2g-2}.$$
Moreover in the case of the conifold,
the function $f(\lambda)$ was related in \nareta\
to the function computed by Schwinger for corrections to the effective
action for a charged scalar in the presence of constant $E,B$ fields.
This has the interpretation of a one loop computation, as in \nareta , where
the light charged
wrapped D3 brane goes around the loop.  The existence of $R^2$
(instead of $1$) reflects the fact that this case has two more
D=4 supersymmetries
than the problem considered by Schwinger.
Certain non-perturbative aspects
of this in connection with pair creation have been discussed in \gopv.
In particular, it was argued that the above expansion of $f$ as a power
series in $\lambda$ should be viewed as an asymptotic expansion and
that there would be corrections of the form ${\rm exp}(-{1\over \lambda})$.
In fact, such corrections are exactly captured by Schwinger's computation
(with a Euclidean circle instanton giving these kind of corrections).

It is natural to ask if there is a parallel situation for the context
we are considering in this paper.  The obvious guess is
to give vev to $H=h$ fields and consider contributions to the action of the
form
\eqn\vevs{S= \int d^{10} x \sqrt{det_{10} g}~ R^4 f(h)}
where
$$f(h)=\sum_g f_g h^{4g-4}$$
and $f_g R^4 H^{4g-4}$ are the corrections we have considered in this paper.
In this
case, we expect strings to be created (at least virtually) and to give
corrections to $R^4$.
We can turn on different types of $H$'s,
and for a generic choice of vevs for $H_{NS-NS}$
and $H_{R-R}$, we should expect production of all $(p,q)$ strings.

One can ask if $f(h)$ (and its non-perturbative
extensions) can be computed in a similar manner as
was done in the case of Schwinger's problem.
If we could
compute $f(h)$ in a different way from perturbative superstring computations,
as was done in the
case of $R^2 F^{2g-2}$ terms near a conifold,
we would be able to fix
the overall normalization for each $f_g$, which we have not fixed
in this paper. It would also combine our conjectures
for all the different $g$'s into a single conjecture.

For concreteness, let us consider corrections involving $H_{R-R}$ in Type IIB
on $R^{10}$.  If we turn on a constant $H_{R-R}$, say in the $H_{012}=h$
direction, we expect that virtual D-strings would be relevant
for computation of $R^4$ corrections.  In analogy with the conifold
problem, we could consider a limit where the D-string becomes light, which
happens at strong coupling of the Type IIB string.  In this limit, by
the $SL(2,Z)$ symmetry of Type IIB, we can view this from the viewpoint
of the dual D-string which now plays the role of the fundamental string, and
for which $H_{R-R}$ is now mapped to $H_{NS-NS}$.  In other words,
we are back to perturbative Type IIB computation of $R^4$ terms in the
presence of constant $H_{NS-NS}$. This, according to our conjecture,
gets infinitely many contributions at genus 0
(one for each $H^{4g-4}$ term)
and one contribution for each genus ($H^{4g-4}$ correction at genus $g$).
And it is as difficult
as the original problem.  So we see that the stringy analog of Schwinger's
computation seems intrinsically stringy and, unlike the conifold
case, we seem not to find a simpler
problem to map it to.

There is, however, one statement we can make.  The process of
nucleating strings from constant $H_{NS-NS}$ background has
been considered in \ref\horet{H. F. Dowker, J. P. Gauntlett, G. W. Gibbons, G.
T. Horowitz, {\it Nucleation of P-Branes and Fundamental Strings},
Phys. Rev. D53 (1996) 7115, hep-th/9512154.}.
In particular, a Euclidean spherical instanton was constructed
with action proportional to $1/h^2$,
on the basis of which it was concluded that the rate
of production of strings should go as ${\rm exp}(-A/h^2)$.
Even though perturbative corrections to the action were not
considered in \horet, their result combined with our
conjectures implies there should be a function
$f(h_+,h_-,\tau)$ whose asymptotic expansion for small $h_+,h_-$
gives the conjecture we have stated in the introduction, and that
there should be corrections of the form ${\rm exp}(-A/h^2)$
completing this function away from $h\sim 0$.

It would be interesting to develop other ways to compute the
overall coefficients of the $R^4 H^{4g-4}$ terms so as to
sum up this series for different $H$'s.  This should teach
us something non-trivial about the creation of $(p,q)$ strings
in background $H$ fields.

\newsec{Concluding Remarks}

In this paper, we have
conjectured the non-perturbative structure of $R^4 H^{4g-4}$
terms in the Type IIB low-energy effective action. The most important evidence
for our conjecture is agreement with explicit $g$-loop superstring
computations.

In string theory, the NS-NS $b_{\mu\nu}$ two-form usually appears with the
graviton in the combination
$g_{\mu\nu}+b_{\mu\nu}$ .
This suggests that $R^4 H^{4g-4}$ terms might be related
by supersymmetry to terms such as $\nabla^{4g-4} R^4$ and $R^{2g+2}$ which
contain the
same number of derivatives but are composed only of graviton fields.
If related,
our conjecture would imply that all such terms are also multiplied by
the Eisenstein
function $E_{g+\half} (\tau)$ in the Type IIB low-energy effective action.
This might relate
our conjecture with at least two other conjectures in the literature.

Based on the known
tree-level $\nabla^{4g-4} R^4$ term in the Type IIB effective action,
Russo conjectured that the non-perturbative $\nabla^{4g-4} R^4$ is multiplied
by precisely the same $E_{g+\half} (\tau)$ function. This is gratifying since
support for
his conjecture (tree-level computations)
comes from a completely different source than the support
for our conjecture ($g$-loop computations).

Also, if $R^{2g+2}$ and $R^4 H^{4g-4}$ terms are
related by supersymmetry,
our results may
be useful for understanding certain M(atrix) model computations.
These
M(atrix) model computations suggest that the effective action of the open
superstring contains $F^{2g+2}$ terms which can be `topologically' computed
at genus $g$ since they contain trivial $\a'$ dependence. Since the open
superstring vertex operator for $F_{\mu\nu}$ is the ``square-root'' of the
Type IIB vertex operator for $R_{\mu\nu\rho\sigma}$, it seems plausible that
the topological
nature of $g$-loop open superstring $F^{2g+2}$ terms are related to the
topological nature of $g$-loop Type IIB $R^{2g+2}$ terms.

\vskip 10pt
{\bf Acknowledgements:}
We would like to thank Michael Green, Hirosi Ooguri, Jorge Russo and
Andrew Strominger for useful discussions. NB would also like to
thank Harvard University for its hospitality and CNPq grant 300256/94-9
for partial financial support.  The research of CV was supported in part
by NSF grant PHY-92-18167.

\listrefs

\end